\documentclass[final,twocolumn,times]{elsarticle}

\usepackage[T1]{fontenc}
\usepackage[utf8]{inputenc}
\usepackage[english]{babel}

\usepackage{graphicx}
\usepackage{dcolumn}
\usepackage{bm}
\usepackage[version=3]{mhchem}
\usepackage{color}
\usepackage[usenames,dvipsnames]{xcolor} 
\usepackage[colorlinks=true,linkcolor=blue,citecolor=blue,urlcolor=blue]{hyperref} 

\usepackage{csquotes}
\usepackage{siunitx}
\usepackage{comment}

\newcommand{\kv}{\mathbf{k}}

\begin{document}

\begin{frontmatter}

\title{From pseudo-direct hexagonal germanium to direct silicon-germanium alloys}%

\author{Pedro Borlido}
\author{Jens Renè Suckert}
\author{Jürgen Furthm\"uller}
\author{Friedhelm Bechstedt}
\author{Silvana Botti}
\author{Claudia R\"odl}
\address{%
Institut f\"ur Festk\"orpertheorie und -optik,
Friedrich-Schiller-Universit\"at Jena, Max-Wien-Platz~1, 07743~Jena, Germany and European Theoretical Spectroscopy Facility
}%

\begin{abstract}
We present \emph{ab initio} calculations of the electronic and optical properties of hexagonal Si$_x$Ge$_{1-x}$ alloys in the lonsdaleite structure. Lattice constants and electronic band structures in excellent agreement with experiment are obtained using density-functional theory. Hexagonal Si has an indirect band gap, while hexagonal Ge has a pseudo-direct gap, i.e.\ the optical transitions at the minimum direct band gap are very weak. The pseudo-direct character of pure hexagonal Ge is efficiently lifted by alloying. Already for a small admixture of Si, symmetry reduction enhances the oscillator strength of the lowest direct optical transitions. The band gap is direct for a Si content below 45\,\%. We validate lonsdaleite group-IV alloys to be efficient optical emitters, suitable for integrated optoelectronic applications.
\end{abstract}


\end{frontmatter}

\section{Introduction}
Alloys of silicon (Si) and germanium (Ge) in the diamond crystal structure with a molar fraction $x$ of Si and $(1-x)$ of Ge, i.e.,
Si$_x$Ge$_{1-x}$, are commonly used as semiconductor materials in integrated circuits, as heterojunction bipolar transistors, or as strained layers in CMOS transistors~\cite{cressler2003}. These Si$_x$Ge$_{1-x}$ alloys are indirect
semiconductors and, hence, not optimal for use in active optoelectronic devices, such as photonic integrated chips~\cite{atabaki2018integrating}. Therefore, over the past years, a lot of effort has been devoted to engineer group-IV materials (and their alloys) with direct gaps and strong dipole-active optical transitions at the minimum band gap 
by straining, nanostructuring or amorphization~\cite{ball2001,vincent_etal_2014,hauge_etal_2015,Assali2019,Ikaros2017,fadalyNature2020}.

At present, the hexagonal (hex) lonsdaleite phase $P6_3/mmc$ ($D^4_{6h}$) of group-IV semiconductors is attracting growing attention. The backfolding of the conduction-band minima of the cubic phase from the $L$ points onto the $\Gamma$ point of the hexagonal Brillouin zone renders hex-Ge a direct semiconductor \cite{rodlPhys.Rev.Materials2019}. However, this direct band gap has a pseudo-direct character, as optical transitions at the lowest gap are very weak. Hex-Si, instead, remains an indirect semiconductor~\cite{rodlPhys.Rev.B2015,rodlPhys.Rev.Materials2019}, since another band minimum at the $M$ point is situated at lower energy than the backfolded minimum at $\Gamma$. Lonsdaleite group-IV materials have been fabricated using ultraviolet laser ablation~\cite{bandetJ.Phys.D:Appl.Phys.2002,haberlPhys.Rev.B2014} or the crystal-phase transfer method~\cite{haugeNanoLett.2015,moran-lopez1992}. Recently, hex-Si and hex-Ge have been realized growing on templates of wurtzite GaP nanowires in form of coreshell nanowires~\cite{conesa-bojNanoLett.2015,fadalyNature2020} with high crystal quality~\cite{fadalyNanoletters2021}. 

The first theoretical study of hex-Si$_x$Ge$_{1-x}$, using a virtual-crystal approximation, suggested a direct-indirect gap crossover for intermediate compositions~\cite{cartoixaNanoLett.2017} and negative direct gaps for Ge-rich alloys. The precise composition $x$ of this crossover and the strength of the lowest-energy optical transitions over the whole composition range remain important open questions which will be addressed in the present work.

\section{Methods}
We have performed density-functional theory (DFT) calculations using the \textsc{VASP} package~\cite{kressePhys.Rev.B1996} with the projector-augmented wave method~\cite{kressePhys.Rev.B1999} and a plane-wave cutoff of 500~eV, treating the shallow Ge~$3d$ electrons as valence electrons. For geometry optimization, we haved employed the PBEsol~\cite{perdewPhys.Rev.Lett.2008} exchange-correlation functional, which gives accurate results for lattice parameters of solids~\cite{csonkaPhys.Rev.B2009,zhangNewJ.Phys.2018}. The Brillouin zone integration was performed using $\Gamma$-centered $\kv$-point grids with a density equivalent to a 12$\times$12$\times$6 mesh for the primitive lonsdaleite unit cell (approximately 5000 points per reciprocal atom~\cite{ongComputationalMaterialsScience2013}). Atomic geometries were relaxed until the forces on the atoms were smaller than 1~meV/{\AA}.

Accurate band structures were computed using the modified Becke-Johnson exchange-correlation potential~\cite{tranPhys.Rev.Lett.2009}, including spin-orbit coupling. Previous work has proven excellent agreement with experiment for the band gaps of semiconductors~\cite{borlidoJ.Chem.TheoryComput.2019} and, in particular, cubic as well as hexagonal Si and Ge~\cite{laubscherJ.Phys.:Condens.Matter2015,rodlPhys.Rev.B2015,rodlPhys.Rev.Materials2019}.

In addition to the band gaps, we also study the radiative lifetime $\tau$ as a global measure of the light-emission properties. To this end, we compute the radiative recombination rate $A_{cv\kv}$ for vertical optical transitions between a conduction state
$\left|c\kv\right>$ and a valence state $\left|v\kv\right>$ with the one-particle energies $\varepsilon_{c\kv}$ and
$\varepsilon_{v\kv}$ as~\cite{deleruePhys.Rev.B1993,dexterSolidStatePhysics1958}
\begin{equation}
   A_{ c v \kv }
   =
   n_\text{eff} \frac{ e^2 (\varepsilon_{c\kv} - \varepsilon_{v\kv}) }{ \pi \epsilon_0 \hbar^2 m^2 c^3 }
   \frac{1}{3} \sum_{j=1}^3 | \langle c\kv | p_j | v\kv\rangle |^2 \, ,
   \label{eq:radiative_recombination_rate}
\end{equation}
with $n_\text{eff}$ the effective refractive index of the medium. The squares of the momentum matrix elements $\langle c\kv | p_j | v\kv\rangle$ are averaged over the Cartesian directions, corresponding to the emission of unpolarized light. The decay rate $1/\tau$ can then be approximated as the thermal average of the recombination rates at temperature $T$,  
\begin{equation}
   \frac{1}{\tau}
   =
   \sum_{cv\kv} A_{cv\kv}\,
   \frac{ w_\kv\, e^{ - (\varepsilon_{c\kv} - \varepsilon_{v\kv})/(k_\text{B} T ) } }
   { \sum\limits_{ c'v'\kv'} w_{\kv'}\, e^{ - (\varepsilon_{c'\kv'} - \varepsilon_{v'\kv'})/(k_\text{B} T ) } } 
   \label{eq:decay_rate}
\end{equation}
with the $\kv$-point weights $w_\kv$.
The direction-averaged refractive indices of hex-Si and hex-Ge are $3.2$ and $3.7$. As in cubic Si$_x$Ge$_{1-x}$ alloys~\cite{jellisonOpticalMaterials1993}, a smooth, but not linear variation can be expected for intermediate compositions. For not introducing a bias in absence of data for the intermediate compositions, we plot $1/(\tau n_\text{eff})$.

We describe 
Si$_x$Ge$_{1-x}$ alloys by a cluster expansion within the strict-regular-solution (SRS) model~\cite{chen1995,sherPhys.Rev.B1987}. Using the \textsc{genstr} tool of the \textsc{ATAT} package~\cite{vandewalleCalphad2002}, we constructed all ordered Si$_x$Ge$_{1-x}$ structures representable by 8-atom supercells (clusters) resulting in 118 symmetry-inequivalent clusters of 3 different cell shapes. Any thermodynamic property $\langle D \rangle$(x) of the alloy can then be written as a weighted average over the cluster properties $D_j$. Within the SRS model, the alloy average translates into
$\langle D \rangle(x) = \sum_j x^0_j\, D_j$, 
with the weights
$x^0_j = g_j ( 1 - x )^{n-n_j} x^{n_j}$ 
of an ideal random alloy. Here, $n$ is the total number of atoms and $n_j$ the number of Si atoms in the cluster cell $j$. The number of symmetry-equivalent atomic configurations represented by each cluster is given by the degeneracy $g_j$. We found cluster sizes of $n=8$ atoms sufficient for converged alloy properties. 

\section{Results and Discussion}

\subsection{Atomic structure}

\begin{figure}
   \centering
   \includegraphics[width=\columnwidth]{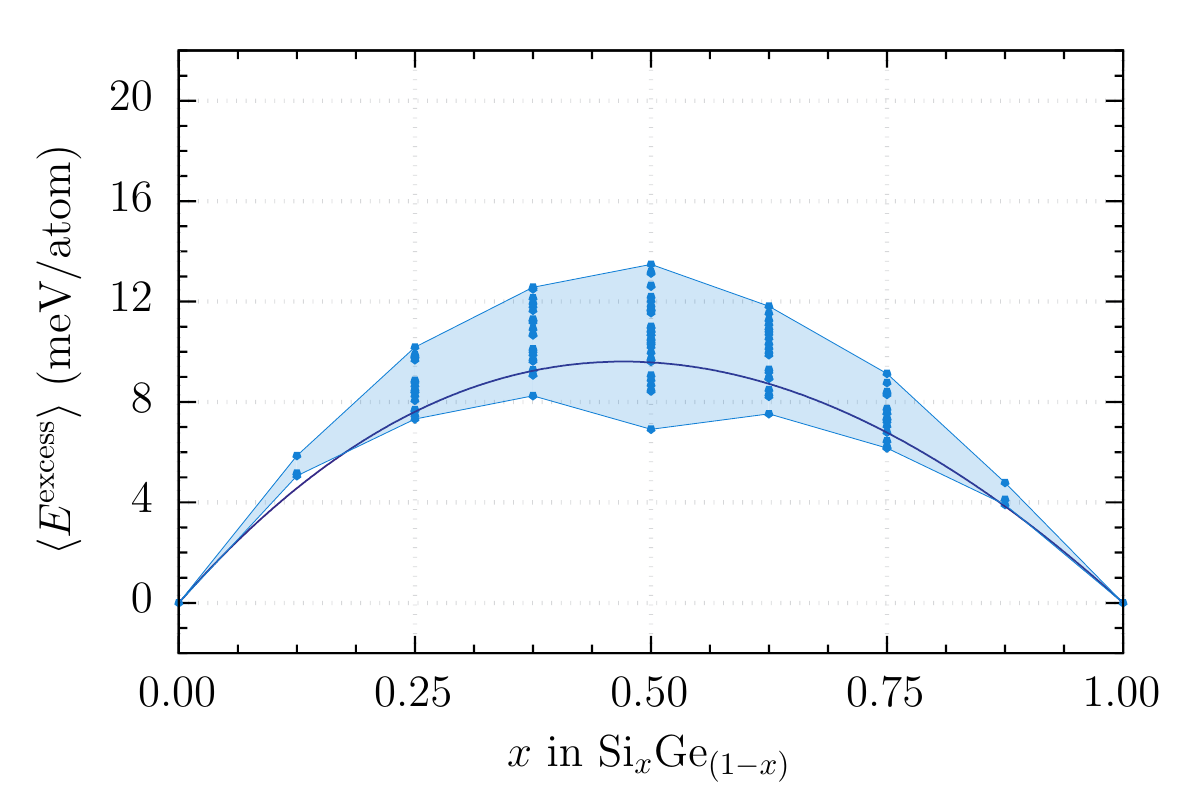}
   \caption{Alloy-averaged excess energy $\langle E^\text{excess}\rangle(x)$ of hex-Si$_x$Ge$_{1-x}$ as a function of composition (solid line). Dots indicate the excess energies $E^\text{excess}_j$ of the individual clusters, with the spread of values at each composition highlighted by the shaded region.}
   \label{fig:energy}
\end{figure}

In Fig.~\ref{fig:energy}, we show the calculated cluster excess energies 
\begin{equation}
E_j^\text{excess} = E_j - \frac{n_j}{n}\, E_\text{Si} - \left(1-\frac{n_j}{n}\right)  E_\text{Ge},
\end{equation}
where $E_j$ is the total energy of cluster $j$ and $E_\text{Si}$ and $E_\text{Ge}$ are the total energies of the end components, for the 8-atom cluster cells. What is more, also the alloy-averaged excess energy $\langle E^\text{excess}\rangle(x)$ is plotted. All cluster excess energies are positive, as the creation of the Si-Ge bond requires energy. The largest spread of values occurs for clusters with stoichiometry $n_j/n=0.5$, ranging from $7$~meV/atom to $14$~meV/atom. The distribution of the excess energies is slightly asymmetric, increasing faster in the Ge-rich region, and it reaches a maximum of $9.6$~meV/atom at $x=0.47$, which is well below the thermal energy at room temperature. This implies that hex-Si$_x$Ge$_{1-x}$ can be treated as ideal random alloy at room temperature, and the SRS model is a justified approximation under these conditions. At lower temperatures, when thermal energies become comparable to the excess energies, more sophisticated alloy statistics, such as the generalized quasichemical approximation~\cite{chen1995,sherPhys.Rev.B1987}, are necessary.

\begin{figure}
\centering
\includegraphics[width=\linewidth]{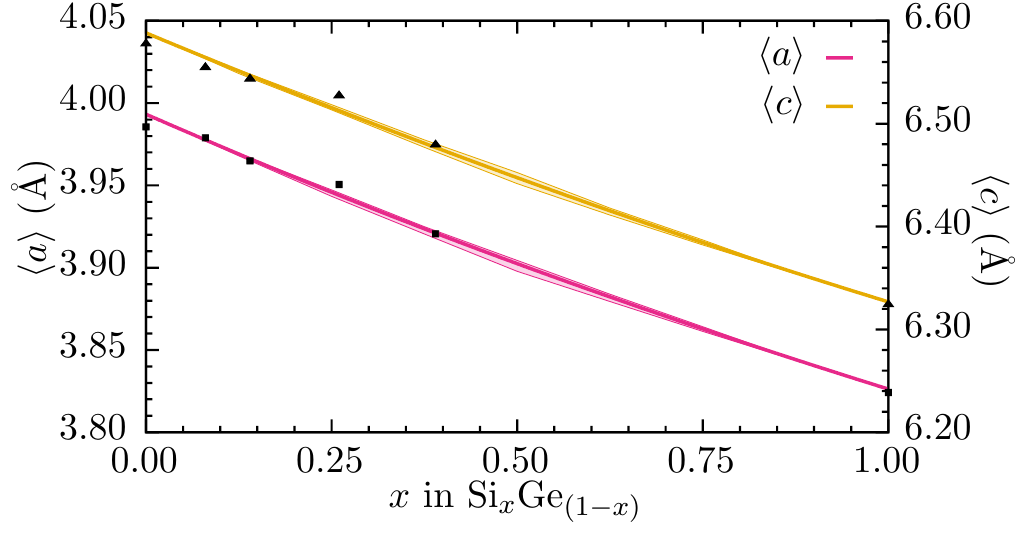} 
\caption{Lattice parameters $\langle a\rangle(x)$  and $\langle c\rangle(x)$ of hex-Si$_x$Ge$_{1-x}$ alloys. The very narrow shaded regions indicate the range of lattice constants obtained for the individual clusters at a given composition. Solid squares and triangles represent experimental values~\cite{haugeNanoLett.2015,fadalyNature2020} for $a$ and $c$, respectively. Experimental error bars are smaller than the used symbols. 
 } 
 \label{fig:average_lattice_const}
\end{figure}

The lattice constants $a$ and $c$ of hex-Si$_x$Ge$_{1-x}$ are shown in Fig.~\ref{fig:average_lattice_const}. They vary almost linearly between $a=3.993$~{\AA} ($c=6.588$~{\AA}) for pure Ge and $a=3.826$~{\AA} ($c=6.327$~{\AA}) for pure Si, which is in agreement with Vegard's law~\cite{vegardZ.Physik1921}. The alloy-averaged lattice parameters $(D=a,c)$ can be fitted to parabolas
\begin{equation}
\langle D\rangle(x) = x D_\text{Si}+(1-x)D_\text{Ge}-x(1-x)b_D,
\end{equation}
with small bowing parameters $b_a = 0.029$~{\AA} and  $b_c = 0.040$~{\AA}. The computed values are in excellent agreement with experimental data from x-ray diffraction~\cite{fadalyNature2020,haugeNanoLett.2017}.

\subsection{Electronic Structure}

\begin{figure*}[t]
   \centering
   \begin{tabular}{cc}
   (a) $n_j/n=0.25$ & (b) $n_j/n=0.5$ \\
   \includegraphics[width=\columnwidth]{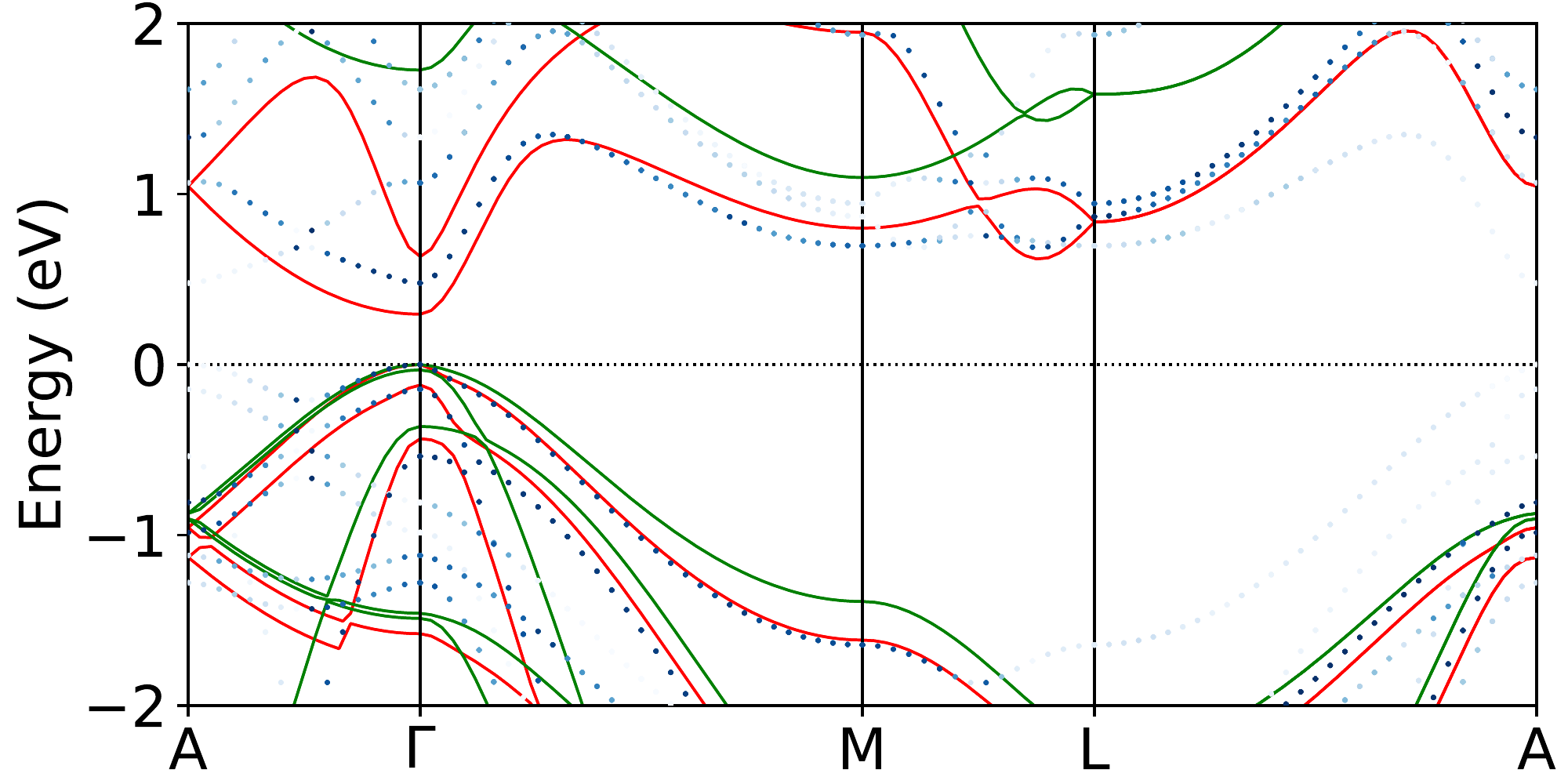} &
   \includegraphics[width=\columnwidth]{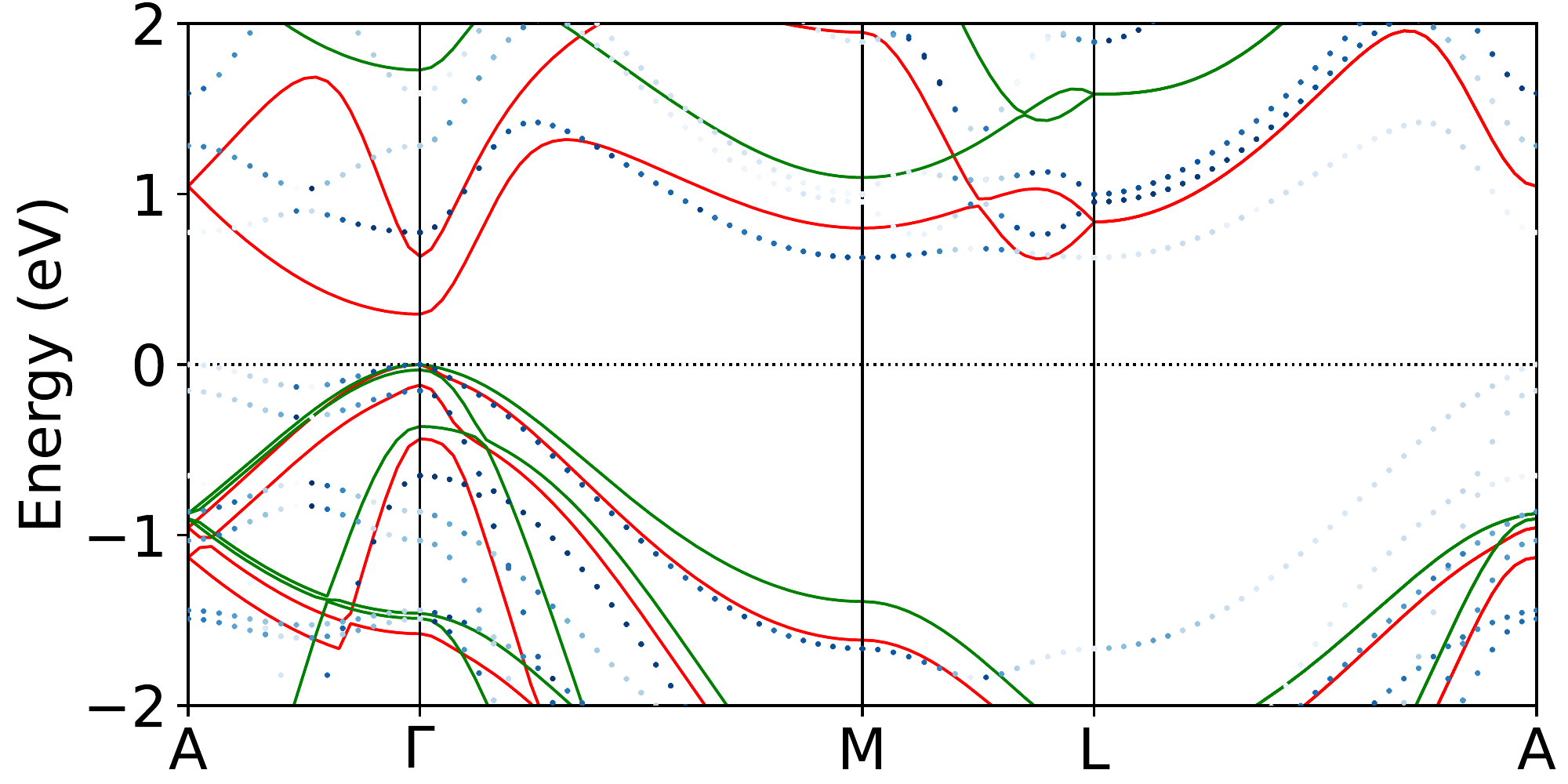}
   \end{tabular}
   \caption{
   Band structures of hex-Si$_x$Ge$_{1-x}$ clusters with $n_j/n=0.25$ and $n_j/n=0.5$. We plot the unfolded band structure of the 8-atom cluster cell with the lowest gap (blue dots) for each composition, as well as the band structures of hex-Si (green) and hex-Ge (red) as a guide to the eye.}
   \label{fig:bands-unfolded}
\end{figure*}

For optoelectronic applications, the light-emission properties of the hex-Si$_x$Ge$_{1-x}$ are of utmost importance. As excited electrons will accumulate in the lowest-energy valleys of the alloy, the clusters with the smallest gap at a given composition are particularly relevant in this context. In Fig.~\ref{fig:bands-unfolded}, the unfolded band structures of the lowest-gap clusters with stoichiometries $n_j/n=0.25$ and $n_j/n=0.5$ are shown along with the band structures of hex-Si and hex-Ge. The 8-atom cluster cells were unfolded onto the lonsdaleite Brillouin zone using \textsc{fold2bloch}~\cite{rubelPhys.Rev.B2014}. The intensity of the blue tone is proportional to the Bloch spectral weight of each state~\cite{rubelPhys.Rev.B2014,wangPhys.Rev.Lett.1998}. It is clearly visible that the fundamental gap is direct for the lowest-gap clusters with $n_j/n=0.25$. The lowest-gap cluster with $n_j/n=0.5$ has an indirect gap. The overall dispersion of the bands remains largely unaffected by stoichiometry and individual bands can still be identified, despite the disorder effects due to alloying. The valence-band maximum is located at the $\Gamma$ point for all clusters. The conduction-band minimum at $\Gamma$, on the other hand, is very sensitive to the composition of the alloy and shifts over a comparably wide range of values, opening the way to tune the band gap of hex-Si$_x$Ge$_{1-x}$ via composition engineering.

\begin{figure}[h]
   \centering
   \includegraphics[width=\linewidth]{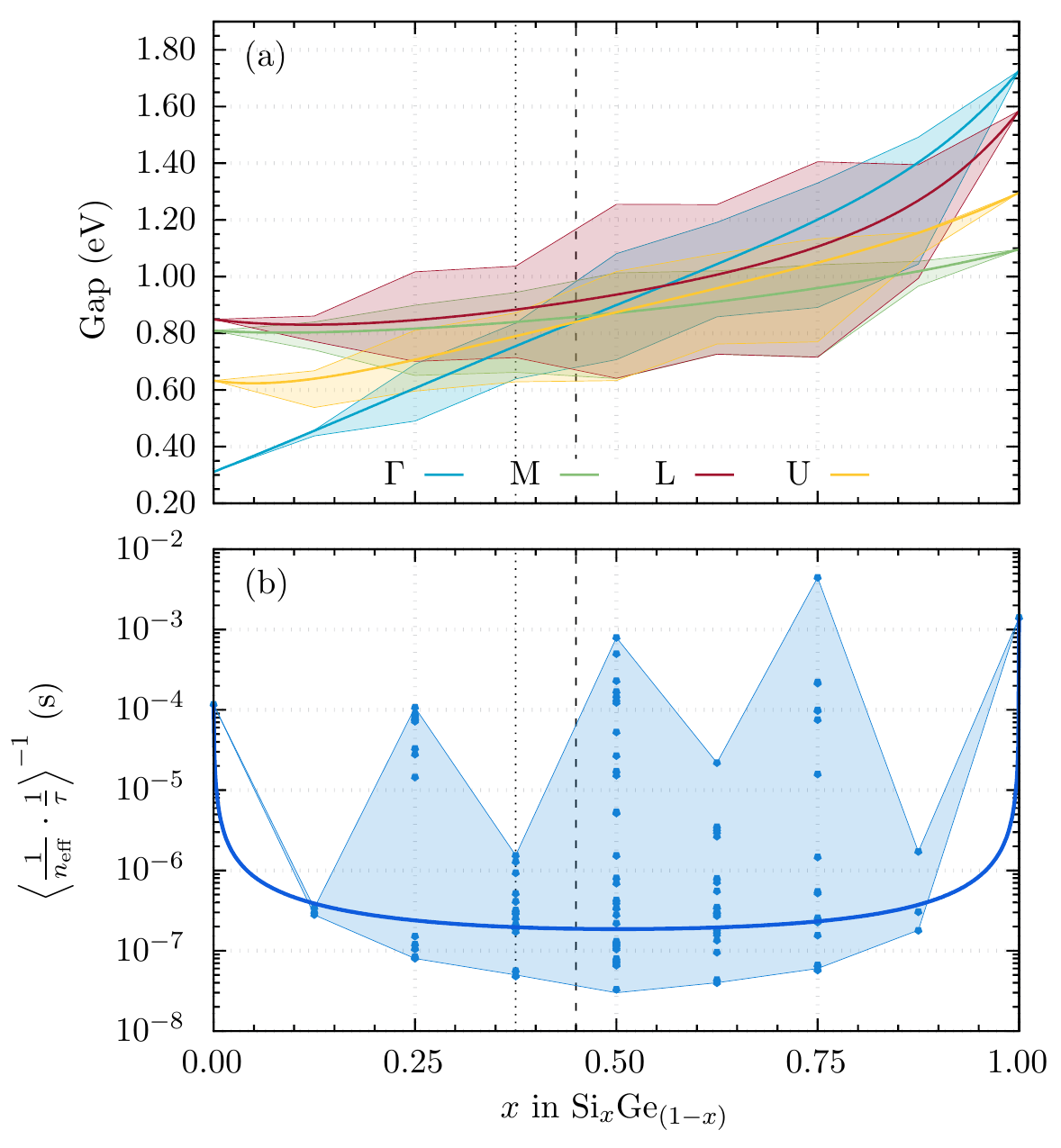}
   \caption{
   (a) Alloy-averaged band gaps as a function of composition.
   Shaded regions show the range of gap values obtained for the individual clusters at a given composition. Vertical dashed (dotted) lines indicate the direct-to-indirect gap transition for the random alloy (the lowest-gap clusters).
   (b) Radiative lifetime as a function of composition (solid line). Dots indicate the results for the individual clusters, with the range at each composition highlighted by the shaded region. 
   }
   \label{fig:avg-gaps}
\end{figure}

In Fig.~\ref{fig:avg-gaps}(a), we trace the evolution of the alloy-averaged conduction-band minima at the most relevant high-symmetry points $\Gamma$, $M$, $L$ and $U$ (which is close to $2/3$ of the $\overline{ML}$ line~\cite{setyawanComputationalMaterialsScience2010,rodlPhys.Rev.Materials2019}) from hex-Ge ($\Gamma$-$\Gamma$ gap of 0.3~eV) to hex-Si ($\Gamma$-$\Gamma$ gap of 1.6~eV, $\Gamma$-$M$ gap of 1.1 eV). The alloy-averaged gap increases with increasing $x$ and is direct for the Ge-rich compositions with $x<0.45$.  For larger $x$, an indirect $\Gamma$-$U$ or $\Gamma$-$M$ fundamental gap appears, depending on the composition. At the direct-to-indirect transition, the gap is about $0.85$~eV. If, in view of discussing emission properties, we consider the lowest-gap clusters at a given composition instead of the alloy average, the direct-to-indirect transition already takes place at $x\approx0.375$, at a gap energy of about $0.63$~eV.

\subsection{Radiative Lifetimes}
To provide a global measure for the light-emission efficiency of hex-Si$_x$Ge$_{1-x}$ alloys, we calculate the alloy average $\langle 1/\tau\rangle^{-1}(x)$ of the radiative lifetime from Eq.~\eqref{eq:decay_rate} at $T=300$~K (see Fig.~\ref{fig:avg-gaps}(b)). Hex-Ge is a pseudo-direct semiconductor, as optical transitions between the top valence and the lowest conduction band are very weak~\cite{suckertPhysRevMaterials2020,rodlPhys.Rev.Materials2019,troncPhys.StatusSolidiB1999}, resulting in long radiative lifetimes (about $10^{-4}$~s at $300$~K). However, transitions involving the second conduction band at about $0.6$~eV from the top valence have  much higher oscillator strengths, comparable to direct III-V semiconductors such as GaAs~\cite{fadalyNature2020}. It has been predicted that moderate tensile uniaxial strain inverts the conduction-band ordering and reduces the radiative lifetime of hex-Ge by more than 3~orders of magnitude~\cite{suckertPhysRevMaterials2020}. 

In hex-Si$_x$Ge$_{1-x}$ alloys, disorder reduces the crystal symmetry such that high light-emission efficiency is possible even without straining the material, since the lowest $\Gamma$-$\Gamma$ transition becomes strongly dipole active for some clusters within the alloy.  As is evident from Fig.~\ref{fig:avg-gaps}(b), this results in an alloy-averaged radiative lifetime that is, throughout the entire range of compositions, 3~orders of magnitude lower than the lifetimes of the end components. As Eq.~\eqref{eq:radiative_recombination_rate} takes only vertical transitions into account, the direct-to-indirect gap transition is not apparent in the lifetimes. However, combining the information on the direct-to-indirect gap transition from  Fig.~\ref{fig:avg-gaps}(a) with the strongly increased optical matrix element for vertical transitions (i.e.\ the by orders of magnitude reduced radiative lifetime) from Fig.~\ref{fig:avg-gaps}(b), we can conclude that hex-Si$_x$Ge$_{1-x}$ alloys in the composition range $0<x<0.4$ are efficient light emitters suitable for optoelectronic applications with tunable band gaps in the spectral range of optical telecommunication.

\section{Summary and Conclusions}
In summary, we performed a comprehensive \emph{ab initio} study of the electronic and optical properties of hex-Si$_x$Ge$_{1-x}$ alloys in view of their light-emission capabilities.
We verified that hex-Si$_x$Ge$_{1-x}$ can be described as ideal random alloy at room temperature and that the lattice parameters closely obey Vegard's law. 
We have shown that Ge-rich hex-Si$_x$Ge$_{1-x}$ alloys can be efficient optical emitters, in contrast to pure hex-Si or hex-Ge. Alloying breaks the crystal symmetry and transforms the pseudo-direct gap of hex-Ge into a strongly dipole active direct gap for the Ge-rich compositions. Overall, the electronic properties of hexagonal group-IV alloys make them very promising for active optoelectronic applications, and will surely keep attracting a fair amount of research in the near future.

\section*{Declaration of Competing Interest}
The authors declare that they have no known competing financial interests or personal relationships that could have appeared to influence the work reported in this paper.

\section*{Acknowledgements}
We acknowledge support from the EU through the H2020-FETOpen project SiLAS (grant agreement No.~735008). C.\,R.\ acknowledges support from the Marie Sk\l{}odowska-Curie Actions (grant agreement No.~751823). Computing time was granted by the Leibniz Centre on SuperMUC (No.~pr62ja). 

\bibliographystyle{elsarticle-num}

\end{document}